\documentclass[12pt]{article} 
\pdfoutput=1 

\usepackage{amssymb}
\usepackage{amsmath}
\usepackage[pdftex]{graphicx}
\graphicspath{{img/}}
\usepackage{epstopdf} 
\usepackage{amsfonts}
\usepackage{bbold}
\usepackage{color}
\usepackage{hyperref}

\newcommand{\eps}{\epsilon}
\newcommand{\rmi}{\mathrm{i}}

\usepackage{siunitx}
\DeclareSIUnit\mT{\milli\tesla}
\DeclareSIUnit\T{\tesla}

\renewcommand{\Im}{\mathop{\mathrm{Im}}\nolimits}

\newenvironment{sciabstract}{%
	\begin{quote} \bf}
	{\end{quote}}

\usepackage[style=nature, backend=bibtex]{biblatex}
\defbibheading{bibliography}{\paragraph{References}}

\usepackage{caption}
\title{Transverse Magneto-Optical Kerr Effect at Narrow Optical Resonances}

\author{O.V.~Borovkova,$^{1,\ast}$ F.~Spitzer,$^{2}$ V.I.~Belotelov$^{1,3}$, I.A.~Akimov$^{2,4,\ast}$, \\
	A.N.~Poddubny$^{4}$, G.~Karczewski$^{5}$, M. Wiater$^{5}$, T. Wojtowicz$^{6}$, \\
	A.K.~Zvezdin$^{1}$, D.R.~Yakovlev$^{2,4}$ and M.~Bayer$^{2,4}$
	\\
	\\
	\normalsize{$^{1}$Russian Quantum Center, 143025 Skolkovo, Moscow Region, Russia}\\
	\normalsize{$^{2}$Experimentelle Physik 2, Technische Universit\"at Dortmund,}\\
	\normalsize{44221 Dortmund, Germany}\\
	\normalsize{$^{3}$Moscow State University, 119991 Moscow, Russia}\\
	\normalsize{$^{4}$Ioffe Institute, Russian Academy of Sciences, 194021 St. Petersburg, Russia}\\
	\normalsize{$^{5}$Institute of Physics, Polish Academy of Sciences, 02668 Warsaw, Poland}\\
	\normalsize{$^{6}$International Research Centre MagTop, Institute of Physics,}\\ 
	\normalsize{Polish Academy of Sciences, 02668 Warsaw, Poland}\\
	\\
	\normalsize{$^\ast$To whom correspondence should be addressed;}\\
	\normalsize{E-mail: o.borovkova@rqc.ru, ilja.akimov@tu-dortmund.de}
}

\date{}

\begin{document}
%
	
	
	\baselineskip24pt
	
	
	\maketitle
	

\begin{sciabstract}
Magneto-optical spectroscopy based on the transverse magneto-optical Kerr effect (TMOKE) is a sensitive method for investigation of magnetically-ordered media.
However, in magnetic materials the optical transitions are usually characterized by spectrally broad resonances with widths considerably exceeding the Zeeman splitting in the magnetic field.
Here we investigate experimentally and theoretically the TMOKE in the vicinity of relatively narrow optical resonances provided by confined quantum systems.
For experimental demonstration we use the exciton resonance in a (Cd,Mn)Te diluted magnetic semiconductor quantum well, where the strong exchange interaction with magnetic ions enables the giant Zeeman splitting of exciton spin states $\Delta$ in magnetic fields of a few Tesla.
In the weak coupling regime, when the splitting $\Delta$ is smaller than the spectral broadening of the optical transitions $\Gamma$, the TMOKE magnitude grows linearly with the increase of the Zeeman splitting and its spectrum has an S-shape, which remains virtually unchanged in this range.
In the strong coupling regime ($\Delta>\Gamma$) the TMOKE magnitude saturates, while its spectrum is strongly modified resulting in the appearance of two separate peaks. The TMOKE is sensitive not only to the sample surface but can be used to probe the confined electronic states in depth if the upper layer is sufficiently transparent.
Our results demonstrate that TMOKE of spectrally narrow resonances serves as a versatile tool for probing the charge and spin structure of electronic states in various confined quantum systems and can be used for spin tomography in combination with the conventional polar Kerr effect.
\end{sciabstract}

\maketitle
\newpage
\noindent
The transverse magneto-optical Kerr effect (TMOKE) is one of the magneto-optical effects that emerge due to the Zeeman splitting of the electron levels in magnetic field~\cite{Zvezdin}. Contrary to the widely used Faraday effect, the TMOKE is sensitive to the spin component orthogonal to the light propagation direction and is usually used to detect the in-plane magnetization of para- and ferromagnetic materials~\cite{Soldatov,MagneticDomains-book,MagneticMicroscopy-book,InPlaneKalish}.

Furthermore, TMOKE is an intensity effect of an incident light beam that is defined by the relative change of reflectance $R$, for the two opposite directions of an in-plane magnetic field \textbf{B}:
\begin{equation}
\label{deltaDef}
\delta=2\frac{R(\mathbf{B})-R(-\mathbf{B})}{R(\mathbf{B})+R(-\mathbf{B})}.
\end{equation}

\noindent TMOKE appears only for oblique light incidence with the magnetic field orthogonal to the incidence plane and in a frequency range where the medium's absorption is non-vanishing~\cite{Zvezdin}. As TMOKE is related to the modification of boundary conditions for the incident light by the magnetic field, it is valuable for investigation of magnetic properties near the sample's interface~\cite{Soldatov,MagneticDomains-book}. If the boundary conditions at two interfaces of a thin magnetic film are different, the transmitted light intensity becomes magnetization dependent as well~\cite{Bonod:04}.

Usually, TMOKE is observed for ferromagnetic metals in the visible and infrared spectral range, where it has a typical value of $\delta \approx 10^{-3}$ and a weak spectral dependence~\cite{Krinchik}. In dielectric materials the off-resonant TMOKE in the transparency range is very weak due to the negligible absorption. However, resonant TMOKE with notable magnitude may be present even in nearly transparent media in the vicinity of the absorption band which is provided by spectrally narrow optical resonances. Such phenomenon has been recently actively studied in hybrid magneto-photonic structures where the optical resonances are represented by electromagnetic modes such as surface plasmon polaritons~\cite{Belotelov2011, Borovkova2018}, waveguide modes~\cite{Chekhov2014, Sylgacheva2016} or optical Tamm states~\cite{TammStates}. It was demonstrated that the TMOKE is drastically enhanced in hybrid metal-dielectric structures at the resonance of surface plasmon-polaritons~\cite{Belotelov2011, Borovkova2016, Bossini2016, Ferreiro2011,Ignatyeva2016} and reaches up to $\delta=0.15$ (see Ref.~\autocite{Pohl}).
Nevertheless, due to the strong absorption in metals the bandwidth of the plasmonic resonances $\Gamma$ significantly exceeds the Zeeman splitting $\Delta$, so that the ratio $\Delta / \Gamma \ll 1$ holds.
In this case the spectral dependence of TMOKE has a characteristic S-shape around the plasmonic resonance: it possesses maxima and minima at the slopes of the resonance and crosses zero at the central frequency.

Further types of optical resonances are represented, for instance, by confined quantum states such as rare earth ions in dielectrics, or excitons in semiconductors. These optical resonances are much narrower as compared to plasmonic excitations. Therefore, the fundamentally new strong coupling regime with $\Delta / \Gamma \geq 1$ can be established, where the exciton damping $\Gamma$ is slower as compared to its spin precession in a magnetic field. Consequently the Zeeman splitting is observed in reflectivity, in contrast to the weak coupling regime with $\Delta / \Gamma \leq 1$.
Surprisingly, TMOKE was not yet studied in this regime irrespective of the studied system.

In this work we reveal TMOKE in the vicinity of narrow optical resonances which originate from excitons. For demonstration purposes we focus on excitons in diluted magnetic semiconductor (DMS) quantum well (QW) structures. Here the exchange interaction of magnetic ions with the conduction band electrons ($s-d$) and valence band holes ($p-d$) allows to achieve a large exciton Zeeman splitting with $\Delta/\Gamma > 1$ already in moderate magnetic fields~\cite{Gaj, Furdyna, DMSbook-Zeeman}. In particular, we demonstrate the enhancement of TMOKE in the vicinity of exciton resonances in a $10\,$nm-thick (Cd,Mn)Te/(Cd,Mg)Te QW.

In what follows, first, we consider the TMOKE theoretically for a generic three-level system and confirm that the TMOKE features observed at the exciton resonance are inherent to a broad class of optical resonances caused by confined quantum states. Second, we demonstrate resonant TMOKE for spectrally narrow excitonic resonances in a (Cd,Mn)Te/(Cd,Mg)Te QW structure.

{\setlength{\parindent}{0cm}
\textbf{Results}\\
\textbf{Theory of TMOKE near the resonance of a three-level system.}
In order to identify the general properties of the TMOKE in the vicinity of a spectrally narrow optical resonance let us theoretically study the TMOKE in a simplified magneto-optical model for the electric dipole optical transitions in a three level quantum system, which is realized in a large variety of atomic, molecular and solid-state objects with pseudospin in the excited or ground state~\cite{Feyman1957,LinToCirc1997,Scully}. In particular, one example of such systems is represented by a single level in the ground state (g) which is coupled through the optical field to two excited states (e, with pseudospin projection $S_y=\pm\frac{1}{2}$). The latter are split due to the Zeeman effect. The selection rules for optical transitions dictate that excitation with an electromagnetic wave, which possesses $\sigma^+$ circular polarization in the $xz$-plane, addresses the state with angular momentum projection $S_y=+\frac{1}{2}$ and correspondingly $S_y=-\frac{1}{2}$ for $\sigma^-$ polarization. Due to the Zeeman splitting, $\Delta$, their energies are split from the central one $E_0$ by $\Delta/2$: $E=E_0\pm\Delta/2$ (see Fig.~\ref{Scheme}a).

These transitions cause resonances in the permittivity tensor $\epsilon_{ij}= \epsilon \delta_{ij}-\rmi e_{ijk} g_k$, where $g_k$ is the gyration vector, $e_{ijk}$ is the Levi-Civita tensor, and $i,j,k=x,y,z$~\cite{Zvezdin}. In the case of the magnetic field directed along the $y$-axis the permittivity tensor components are:

\begin{equation}
\label{Eq:epsil}
\begin{split}
\epsilon & = \epsilon_\text{b}\left(1+\frac{4\pi |\mathbf{d}|^2}{\epsilon_\text{b}} \frac{E_0- E - \rmi \Gamma}{(E_0- E - \rmi \Gamma)^2 -(\Delta/2)^2}\right), \\
g_{y} & = \frac{ 2 \pi |\mathbf{d}|^2 \Delta}{(E_0- E - \rmi \Gamma)^2 -(\Delta/2)^2};\;\; g_x=0;\;\; g_z=0,
\end{split}
\end{equation}
where $E$ is the energy, $\epsilon_b$ is the background permittivity, $\mathbf{d}$ is the matrix element of the electron transition dipole moment and $\Gamma$ is the damping.

Relatively simple analytical expressions for TMOKE (Eq.~(S1) in Supplementary) can be obtained for the light reflection from a semi-infinite magnetic medium. The principle scheme of the considered geometry of the light incidence, and the applied magnetic field is shown in Fig.~\ref{Scheme}b.

We start the asymptotic analysis of Eq.~(S1) for the weak coupling regime $(\Delta/\Gamma \ll 1)$ which gives the following expression for TMOKE (see Supplementary, Eq.~(S2)):

\begin{equation}
\label{Eq:Deltasmall}
\delta=-\frac{8 E_d \Delta \Gamma \tan \theta }{\epsilon_\text{b}-1}\frac{\widetilde{E}}{[ (\widetilde{E}+a)^2 + \Gamma^2][(\widetilde{E}-a)^2 + \Gamma^2 ]},
\end{equation}

\noindent where we define the energy related to the dipole moment as $E_d = 4 \pi |\mathbf{d}|^2/\epsilon_\text{b}$, and we introduce $\widetilde{E} = (E_0-E)+E_d(1+\frac{1}{2(\epsilon_\text{b}-1)})$ and $a=\frac{1}{2(\epsilon_\text{b}-1)}E_d$.
Here and below we assume that the incidence angle $\theta$ is small enough so that we can neglect $\tan^2\theta$ with respect to $\epsilon_\text{b}$.
As follows from Eqs.~\eqref{Eq:epsil} and \eqref{Eq:Deltasmall} the TMOKE is proportional to the damping $\Gamma$ and exists only within the spectral range where the imaginary part of the permittivity is not vanishing.
In this sense, it is similar to the effect of magnetic dichroism.
The TMOKE grows linearly with $\Delta$ and its spectrum has an antisymmetric S-shape with respect to $E=E_0+E_d(1+\frac{1}{2(\epsilon_\text{b}-1)})$ (see Fig.~\ref{Profiles}a).
The TMOKE has a maximum and a minimum spectrally separated by $\Gamma$.
The magnitudes of the TMOKE extrema are directly proportional to the Zeeman splitting $\Delta$:
\begin{equation}
\label{Eq:reson1}
\delta=\mp \frac{3 \sqrt{3} E_d \Delta \tan \theta }{2 ( \epsilon_\text{b}-1) \Gamma^2}.
\end{equation}
At the same time the shape of the TMOKE spectrum does not change with the magnetic field for $\Delta/\Gamma \ll 1$.

Narrow optical resonances in sufficiently large magnetic fields lead to the other limiting case when $\Delta/\Gamma \gg 1$. Analysing the general expression for TMOKE (Eq.~(S1)) we should take into account that the resonances of TMOKE are located at $|E_0-E|=\Delta/2$. We treat $\Gamma/\Delta$ as a small parameter and derive the analytical expression for the TMOKE peaks in case $\Delta/\Gamma \gg 1$ (details in Supplementary section A):

\begin{equation}
\label{Eq:largeB}
\delta=\mp \frac{4 \epsilon_\text{b} E_d \tan \theta}{\Gamma} \frac{\epsilon_\text{b}^2-\epsilon_\text{b}- \frac{E_d^2}{4\Gamma^2}}{(\epsilon_\text{b}^2-\epsilon_\text{b}- \frac{E_d^2}{4\Gamma^2})^2+(\frac{2\epsilon_\text{b}-1}{2 \Gamma})^2 E_d^2}
\end{equation}

\noindent Minus refers to $E_0-E>0$, and positive TMOKE corresponds to $E_0-E<0$. TMOKE spectra for the strong coupling regime are presented in Fig.~\ref{Profiles}b. Contrary to the previous situation of weak coupling, the relatively large magnetic field changes the shape of the TMOKE spectrum when $\Delta > \Gamma$: it comprises two separate peaks of opposite sign, the energy distance between the peaks corresponds to the Zeeman splitting $\Delta$ and consequently increases with the applied magnetic field. Another interesting feature of the strong coupling regime is the fact that the TMOKE peaks' magnitude experiences saturation. When $\Delta$ and $\Gamma$ are almost equal, one can see that the TMOKE peaks behave like in the weak coupling regime, and the maximal TMOKE gradually grows. However, this initial increase of TMOKE ceases, when the ratio $\Delta/\Gamma$ exceeds about 5. Now the relatively large Zeeman splitting approximation is fulfilled. Henceforth, the maximum value of TMOKE does not depend on $\Delta$. Similarly to the weak coupling regime, the dependence of TMOKE on the incidence angle is linear as one can see from Eq.~(\ref{Eq:largeB}).

Thus, in large magnetic fields, when the Zeeman splitting exceeds the resonance linewidth, one can observe novel features that do not exist in case of small magnetic fields. Namely, the TMOKE spectrum's shape changes with magnetic field: it has two peaks of opposite sign and these peaks move away from each other with increasing magnetic field. Moreover, on the contrary to the weak coupling regime, the peaks' magnitude does not depend on the applied magnetic field. These general features have been ascertained for the simplest case when the $p$-polarized beam is reflected from a semi-infinite magnetic material with one narrow resonance. For more complex systems additional peculiarities emerge, nevertheless, preserving the general properties discussed above.

As confined quantum states can be formed not only near the sample surface but also at some region in the bulk (e.g. in the QW inside a semiconductor sample with a cap layer), it is important to consider the case when the magneto-optical resonance appears in a thin layer buried at some distance from the sample surface. Though TMOKE is considered as a surface sensitive effect, in this case it can be useful for probing confined resonances in depth. For this purpose we analyse a trilayer structure composed of: the cap layer of varying thickness, the \SI{10}{nm}-thick QW layer with the dielectric permittivity tensor given by Eq.~(\ref{Eq:epsil}) and the buffer layer (see Fig.~\ref{Fig:Sample}a). Variation of the cap layer thickness modifies the shape of the TMOKE resonance with a period of \SI{120}{nm} which is half of the light's wavelength in the cap layer (Fig.~\ref{Profiles}c).
Additionally one can achieve a fully antisymmetric TMOKE spectral shape as it occurs for the \SI{90}{nm}~cap layer.

The TMOKE features discussed above can be observed in different systems that posses sharp optical resonances originating from confined quantum states. In the next section we concentrate on the experimental demonstration of these features for the case of exciton resonances in magnetic semiconductor QW structures.

\textbf{TMOKE at the exciton resonance in semiconductor quantum well.}
To investigate the TMOKE mediated by excitons in diluted magnetic semiconductors, we use a sample with a QW structure grown by molecular-beam epitaxy on a (001) oriented GaAs substrate. The \SI{10}{\nm}-thick DMS Cd$_{0.974}$Mn$_{0.026}$Te QW layer is sandwiched between non magnetic Cd$_{0.73}$Mg$_{0.27}$Te barriers (buffer and cap layers) (see Methods and Fig.~\ref{Fig:Sample}a). The TMOKE measurements were performed in two regimes: (i) The weak coupling regime in magnetic field of $B = \SI{580}{mT}$ at a temperature $T\approx \SI{10}{\kelvin}$; (ii) The strong coupling regime where the magnetic field was varied in \SI{125}{\milli\tesla} steps from \SI{0.5}{} to \SI{5.0}{\tesla} at $T=\SI{2}{\kelvin}$. For the first case the TMOKE spectra were obtained in a wide range of incidence angles using a Fourier imaging setup. In the second case the measurements were performed for $\theta = 5^\circ$. In both cases the magnetic field was oriented in the QW plane and perpendicular to the plane of light incidence as it is shown in Fig.~\ref{Fig:Sample}a. The details on the measurement procedure and evaluation of TMOKE spectra are presented in the Methods section.
\\

\textbf{TMOKE in the weak coupling regime.}\label{TMOKE-lowB}
Figure~\ref{Fig:Exp} summarizes the data obtained in the weak coupling regime of excitonic states. Reflection spectra of the structure show oscillations related to the interference of light after multiple reflections within the (Cd,Mg)Te buffer layer (solid curve in Fig.~\ref{Fig:Exp}a). The exciton resonance from the DMS QW is represented by a weak feature in the reflection spectrum at a photon energy of around \SI{1.683}{\eV}. In contrast, the TMOKE spectrum presented in Fig.~\ref{Fig:Exp}b at $\theta = \SI{5}{\degree}$ shows a much more abundant picture with two resonances located at \SI{1.683}{} and \SI{1.701}{\electronvolt}. This result highlights the sensitivity of TMOKE spectroscopy.

The presence of two exciton resonances in the TMOKE spectrum of the DMS QW structure is related to the quantization of carriers inside the QW as well as uniaxial strain in the direction along the normal to the QW~\cite{Ivchenko2005}. For that reason the light-hole and heavy-hole exciton resonances, which are degenerate in a bulk crystal, split into the light-hole (lh) state with $E_{lh}(B=0) = \SI{1.701}{\electronvolt}$ and the heavy-hole (hh) state with $E_{hh}(B=0)=\SI{1.683}{\electronvolt}$.
The corresponding energies are marked by dotted vertical lines in Figs.~\ref{Fig:Exp}a,b. The light-heavy-hole splitting is equal to $\Delta_{lh}=\SI{18}{\milli\electronvolt}$. Both TMOKE resonances are observed for oblique incidence near the frequencies of the corresponding transitions from the hole states and have antisymmetric S-shape. The exciton mediated TMOKE grows with increase of the incidence angle, reaching 1.1\% at $\theta = 20^\circ$ (Fig.~\ref{Fig:Exp}c), which is several orders of magnitude larger than the off-resonant value of $0.01 \%$, observed at $\theta = \SI{20}{\degree}$ and \SI{1.578}{\electronvolt}. 

We emphasize that the light-hole exciton feature in the photoluminescence and reflectivity spectra is too weak to determine the position of the resonance which is mainly due to its fast relaxation into the ground state corresponding to the heavy-hole exciton. In contrast to that, the TMOKE spectrum can be used to determine the positions of both exciton resonances. Interestingly, the strength of TMOKE is about twice larger for light-hole excitons than for the heavy-hole state and the sign of the effect is opposite. In other words, for positive incidence angles $\theta$ the TMOKE magnitude for the lh exciton changes from positive to negative values with increasing photon energy $E$ (rising S-shape), while for the hh exciton an opposite behaviour with falling S-shape is observed, as shown in Fig.~\ref{Fig:Exp}b.

Heavy-hole excitons show a smaller magnitude of TMOKE because at zero magnetic field the corresponding transitions are linearly polarized in the QW plane and hence carry zero spin along the magnetic field direction. Hence, the TMOKE for the heavy-hole resonance is enabled only due to the admixture of light-hole states at non-zero magnetic field. That explains why the heavy-hole TMOKE remains relatively weak until the Zeeman splitting becomes comparable with the light-heavy-hole splitting. It also hints to the origin of the different sign for light-holes and heavy-holes that will be explained more intuitively once we consider higher magnetic fields.

Now we proceed to a quantitative description of the TMOKE resonance for hybridized light-hole and heavy-hole states at weak magnetic field. To that end we have calculated the effective permittivity tensor of the QW. The orbital effects of the magnetic field, that can mix light-hole and heavy-hole resonances and induce spatial dispersion~\cite{Kotova2018}, were neglected. In the case of a DMS structure these effects are less important than the Mn-mediated giant Zeeman splitting.
The resulting QW permittivity tensor can be presented as
$\eps=\eps_{b}(1+\chi_{hh}+\chi_{lh})$, with the heavy-hole and light-hole susceptibilities in the $xz$ subspace ($Ox$ is in the QW plane and perpendicular to the external field and $Oz$ is normal to the QW plane, see Fig.~\ref{Fig:Sample}a). The susceptibilities read (see section B of Supplementary Materials for the derivation details):
\begin{align}\label{eq:chiSmall}
\chi_{hh}&=\frac{\hbar\omega_{LT}D_{hh}}{
D_{hh,+}D_{hh,-}
}\begin{pmatrix}
\mathcal Z^{2}&\rmi \mathcal Z/3\\-\rmi \mathcal Z/3 & 1
\end{pmatrix}\:,\\
\chi_{lh}&=\frac{\hbar\omega_{LT}D_{lh}}{
D_{lh,+}D_{lh,-}
}\begin{pmatrix}
\frac{4}{3}&-\frac{\rmi}{3}\left(\frac{E_{lh,-}-E_{lh,+}}{D_{lh}}+\mathcal Z\right )\\
\frac{\rmi}{3}\left(\frac{E_{lh,-}-E_{lh,+}}{D_{lh}}+\mathcal Z\right )&\frac{1}{3}
\end{pmatrix}\:.\nonumber
\end{align}
Here, $D_{\nu}\equiv E_{\nu}-E-\rmi\Gamma_{\nu} $; $E_{\nu}$ and $\Gamma_{\nu}$ are the exciton resonance energy and width, the index $\nu=lh,hh,(lh,\pm),(hh,\pm)$ labels zero-field and Zeeman-split excitonic transition energies (see also Fig.~\ref{Fig:Sample}c) and $\omega_{\rm LT}$ is the effective longitudinal-transverse splitting. The parameter $\mathcal Z$
depends linearly on the magnetic field being responsible for the light-heavy-hole mixing. It is determined by the ratio of the Zeeman splitting of the heavy-hole state in the Faraday configuration $\Delta_{h,F}$ to the light-heavy-hole splitting $\Delta_{lh}$. For the experimentally relevant case of $B=\SI{580}{\milli\tesla}$ it is equal to $|\mathcal Z|\approx 0.3$. In the general case one has $\mathcal Z = N_0 \beta x S B_S/\Delta_{lh}$, where ${N_{0}}\beta = \SI{-0.88}{\electronvolt}$ is the $p-d$ exchange interaction constant, $S=5/2$ is the Mn spin, $x$ is the concentration of Mn$^{2+}$ ions and $B_S(\varepsilon)$ is the modified Brillouin function describing the magnetization of Mn ions at given magnetic field and temperature~\cite{Gaj, Furdyna, DMSbook-Zeeman, Landwehr}. For $S=5/2$ the argument of $B_S(\varepsilon)$ is defined as $\varepsilon = 5\mu_B g_{\rm Mn}B / 2k_B T_{\rm eff}$ with the Bohr magneton $\mu_B$, the Boltzmann constant $k_B$, the Land\'e factor of Mn$^{2+}$ ion $g_{\rm Mn} = 2.01$ and the effective temperature $T_{\rm eff} = T + T_0$. The specific parameters of the investigated structure $x \approx 0.026$ and $T_0\approx \SI{1.5}{\kelvin}$ were evaluated from the magnetic field dependence of heavy-hole exciton splitting in the Faraday geometry $\Delta_{h,F}$ measured using photoluminescence spectroscopy~\cite{Spitzer-TMRLE-Nature}.

The TMOKE response is described by the off-diagonal elements of the susceptibilities \eqref{eq:chiSmall}. The different signs before these elements for light-holes and heavy-holes reflect different signs of the TMOKE response. Because of the term $\propto (E_{lh,-}-E_{lh,+})/{D_{lh}}$, for light-holes TMOKE is possible even when the mixing with the heavy-holes is neglected, i.e. $\mathcal Z\to 0$. Note that the mixing term $\mathcal Z$ is determined solely by the energy structure of the valence band states, i.e. proportional to $\Delta_{h,F}/\Delta_{lh}$, while the energy splitting of the exciton resonances is governed by the Zeeman splitting of both valence band and conduction band states as it is shown in Fig.~\ref{Fig:Sample}b,c. The contribution from the electrons is given by $\Delta_{e}=N_0 \alpha x S B_S$, where $N_0\alpha = \SI{0.22}{\electronvolt}$ is the $s-d$ exchange interaction constant between the conduction band electrons and Mn$^{2+}$ ions in (Cd,Mn)Te. The corresponding splittings for light-hole and heavy-hole excitons are given by $\Delta_{X,\nu}=E_{\nu,+}-E_{\nu,-}+\Delta_e$ with $\nu= lh, hh$, respectively. A detailed description of the magnetic field induced splitting of exciton resonances in the Voigt geometry is given in section B of the Supplementary Materials.

Using the transfer matrix method for multilayer DMS structures with the QW described by Eq.~\eqref{eq:chiSmall}, we calculated the reflectance and TMOKE spectra.
They show good agreement with the experimental data if $E_{lh}=\SI{1.7005}{\eV}$ and $E_{hh}=\SI{1.6825}{\eV}$ are assumed (dashed curves in Fig.~\ref{Fig:Exp}a,b). The longitudinal-transverse splitting is $\hbar \omega_{LT}=\SI{0.65}{\meV}$ while
the linewidth and Zeeman splitting of both resonances are $\Gamma_{hh}=\Gamma_{lh}=\SI{2.4}{\meV}$, $\Delta_{X, hh}=\SI{1.2}{\meV}$ and $\Delta_{X, lh}=\SI{4.3}{\meV}$.
Therefore, in the external magnetic field of \SI{580}{\milli\tesla} $\Delta$ is of the order of $\Gamma$, and in Fig.~\ref{Fig:Exp}(b,c) we deal with the weak coupling regime.
\\

\textbf{TMOKE in the strong coupling regime.}
As the exciton resonances are quite narrow the strong coupling regime ($\Delta/\Gamma \gg 1$) is already reachable in magnetic fields of several Tesla, i.e. if the temperature is kept low enough ($T=\SI{2}{\kelvin}$) that the ratio $\Delta/\Gamma$ exceeds $5$ at $B=\SI{2}{\T}$ and is about $10$ at $B=\SI{5}{\T}$.

The TMOKE spectra and their dependence on magnetic field strength up to \SI{5}{\tesla} for $\theta=5^\circ$ are presented in Fig.~\ref{Fig:HighField}. It follows that for $\Delta/\Gamma \gg 1$ each of the resonant TMOKE features corresponding to heavy and light-hole excitons at $B=0$ split into two resonance features. The splitting increases with the increase of $B$. Remarkably, the TMOKE maximum/minimum values are saturated at $B=\SI{1}{\T}$ and remain constant for higher fields. Consequently, in this regime the magnetic field influences the shape of the TMOKE spectrum but does not increase its value of $\delta = \SI{.5}{\percent}$.

In order to interpret the four branches in the TMOKE spectra at higher $B$ let us consider the case of the Zeeman energy which overcomes the light-heavy-hole splitting ($\mathcal Z \gg 1$) in magnetic fields higher than $\SI{1}{\T}$ (see Fig.~\ref{Fig:Sample}). In this regime a simple qualitative picture can be considered which is realized in a bulk (Cd,Mn)Te crystal. In Voigt configuration ($\mathbf B \parallel Oy$) the Zeeman-split electron states are characterized by the spin projection $S_{e,y}=\pm 1/2$ representing a doublet. The valence band states split into a quartet with Zeeman energy governed by the total angular momentum projection on the magnetic field direction $J_{h,y}=\pm 1/2,\pm 3/2$. Note that the case of strong magnetic field is fundamentally different from the case of weak magnetic field where the quantization axis is given by the QW confinement potential and therefore the energies of the valence band states are mainly determined by the normal projection of the angular momentum $J_{h,z}=\pm 1/2,\pm 3/2$ for light-holes and heavy-holes, respectively.

Neglecting the QW confinement the splittings of conduction and valence band states are given by $\Delta_e = 2 N_0 \alpha x \left|S_{e,y}\right| B_S$ and $\Delta_h = -\frac{2}{3} N_0 \beta x \left|J_{e,y}\right| B_S$, respectively.
Thus, we obtain four $\sigma$ polarized exciton transitions. The energy splitting between the outer heavy-like hole components $\Delta_{\rm out}=\Delta_h+\Delta_e$ correspond to the excitonic states $|\pm3/2,\pm1/2\rangle$ written in the $|J_{h,y},S_{e,y}\rangle$ notation.
The inner components are split by $\Delta_{\rm in} = \Delta_h/3-\Delta_e$ corresponding to light-hole-like exciton resonances $|\pm1/2,\mp1/2\rangle$.
As follows from Fig.~\ref{Fig:HighField} the splitting $\Delta_{\rm out}\approx$ \SI{70}{\milli\electronvolt}, while $\Delta_{\rm in}$ is about 8 times smaller at $B=\SI{5}{\T}$.
According to the $s-d$ and $p-d$ exchange constants in bulk (Cd,Mn)Te the relation $\Delta_{\rm out}>\Delta_{\rm in}>0$ holds~\cite{DMSbook-Zeeman}.
In this case the two optical transitions with lower energy $|-3/2,-1/2\rangle$ and $|-1/2,+1/2\rangle $ are $\sigma^+$ polarized in the $xz$-plane, while the upper energy transitions $|+3/2,+1/2\rangle$ and $|+1/2,-1/2\rangle$ have opposite $\sigma^-$ polarization (see also Fig.~\ref{Fig:Sample}c).
The form of the S-shape resonances in large magnetic field is in full accord with this considerations: it has a rising slope for the low energy exciton transitions at \SI{1.666}{} and \SI{1.688}{\eV} and changes to falling for the higher energy resonances at \SI{1.696}{} and \SI{1.733}{\eV}.
This also explains the opposite sign of the light-hole and heavy-hole TMOKE resonances observed in the limit of small magnetic fields (Fig.~\ref{Fig:Exp}).

The susceptibilities at large fields can be simplified to a compact antisymmetric form (Supplementary section B):
\begin{align} \label{eq:chiBig}
\chi_{hh}&=
\frac{\hbar\omega_{LT}}{6}
\begin{pmatrix}
U_{hh}&-\rmi W_{hh}\\
\rmi W_{hh}&U_{hh}\\
\end{pmatrix}\:,\\
\chi_{lh}&=\nonumber
\frac{\hbar\omega_{LT}}{2}
\begin{pmatrix}
U_{lh}&\rmi W_{lh}\\
-\rmi W_{lh}&U_{lh}\\
\end{pmatrix}
\end{align}
where $U_{\nu}=\frac1{D_{\nu,+}}+\frac1{D_{\nu,-}}$ and $W_{\nu}=\frac1{D_{\nu,+}}-\frac1{D_{\nu,-}}$, and $\nu=lh,hh$.
It allows one to calculate the position of the four exciton states versus the magnetic field (dashed curves in Fig.~\ref{Fig:HighField}) and satisfactory describe the experimental TMOKE map for large magnetic fields (Fig.~\ref{Fig:HighField}b). In particular, this model also predicts the saturation of the TMOKE in higher magnetic fields.

{\setlength{\parindent}{0cm}
\paragraph{Discussion}$~~$\\
To conclude, we have demonstrated here the enhancement of TMOKE in the vicinity of a narrow optical resonance. There are two distinct TMOKE regimes which depend on the magnitude of the Zeeman splitting with respect to the linewidth. In the weak coupling regime ($\Delta < \Gamma$) the TMOKE magnitude grows linearly with the increase of the Zeeman splitting and its spectrum has an S-shape.
On the contrary, in the strong coupling regime ($\Delta > \Gamma$) the TMOKE magnitude saturates, while its spectrum is strongly modified with the increase of the Zeeman splitting leading to the appearance of two separate TMOKE peaks.
The spectral dependence of the TMOKE peaks and the sign of $\delta$ provide rich information on the energy structure and selection rules for optical transitions.

Though in this work the considered phenomena were experimentally demonstrated at the exciton resonances in a \SI{10}{\nano\meter}-thick (Cd,Mn)Te/(Cd,Mg)Te QW, they can also be found in other physical systems. For example, the optical resonances of the rare-earth ions in dielectric materials are quite narrow so that $\Gamma \le \Delta$ even in moderate magnetic fields. In this respect, X-ray spectra of the confined quantum states are of interest~\cite{Fukui,Kang,Yagupov2016}. For instance, the rare-earth ions in orthoborates provide lines with $\Delta/\Gamma \approx 1$ in $B \approx \SI{1}{T}$~\cite{Yagupov2016}. Apart from X-ray, the visible and infrared spectra also demonstrate narrow lines~\cite{RE-fiberlaser,Agrawal,Vinh,BOROVIKOVA2015,Popova2009}.

Our results have several implications in fundamental and applied optics. First, resonant TMOKE in the strong coupling regime can be used as a spectroscopic tool for investigation of the energy structure of electronic states which are involved in optical transitions. It should be noted that though TMOKE is usually considered as a surface sensitive effect, it is demonstrated here to be applicable for probing confined resonances in the depth of the sample as well.

Next, resonant TMOKE can be used to read out the in-plane magnetization of the system which is appealing for possible applications in quantum information and sensing technologies~\cite{Awschalom-2010, Dyakonov-book}. Typically, the spin dynamics in semiconductors are assessed by the polarization rotation due to the Faraday effect or Polar Kerr effect (PMOKE)~\cite{Buda-1994}. However, these give access only to one of the spin components parallel to the direction of light propagation ($\mathbf{B}\parallel\mathbf{k}$). On the other hand, the TMOKE allows measuring the perpendicular component of the magnetization. Thus, a combination of TMOKE and PMOKE, provides the possibility for performing spin tomography.

{\setlength{\parindent}{0cm}
\paragraph{Methods}$~~$\\
To investigate the TMOKE mediated by the excitons in a DMS, we used a quantum well structure grown by molecular beam epitaxy on a \SI{400}{\um}-thick (001) oriented GaAs substrate.
The \SI{10}{\nm}-thick magnetic Cd$_{0.974}$Mn$_{0.026}$Te quantum well layer is sandwiched between non-magnetic Cd$_{0.73}$Mg$_{0.27}$Te barriers (\SI{3.25}{\micro\meter} thick buffer and \SI{250}{\nano\meter} thick cap layer, respectively).

The experimental studies were performed on the following two set-ups.
In the weak coupling regime the sample was placed in a He flow cryostat between the ferrite cores of an electromagnet.
Fourier imaging spectroscopy was used to measure the angular- and wavelength- resolved reflectivity and TMOKE spectra at low temperatures of about \SI{10}{\K} using a tungsten halogen lamp, which illuminates the sample with $p$-polarized light.
The reflected light was collimated using a microscope objective with numerical aperture of $0.4$, resulting in the experimentally accessible angular range of $\pm23^\circ$.
A telescope consisting of two achromatic doublets mapped the collimated light onto the spectrometer slit. The exit slit of the spectrometer was equipped with a thermoelectric cooled charge coupled device (CCD) detector, providing a spectral resolution of \SI{0.6}{\nm} and an angular resolution of about \SI{0.4}{\degree}.
By taking the reflected intensity for two opposite magnetic field directions the parameter $\delta$ characterizing the TMOKE was deduced.
For these rather low magnetic field strengths, the noise level is reduced by repetitive switching of the magnetic field direction.
This also smoothes potential fluctuations of the lamps intensity.

On the other hand, in the giant Zeeman splitting case the sample was kept at a temperature of \SI{1.6}{\kelvin} inside a liquid helium bath cryostat equipped with a split-coil superconducting magnet.
Transverse magnetic fields of up to \SI{5}{\tesla} were applied in order to observe the magnetic field induced change of reflectivity for $p$-polarized white light incident under an angle of $5^\circ$.
The reflected light was subsequently dispersed by the spectrometer and detected with the CCD detector.
Sweeping of the current in the superconducting magnet is slow and therefore it does not allow to perform repetitive measurements in two opposite magnetic fields within a reasonable time.
Therefore we sweeped the magnetic field once in steps of \SI{125}{\milli\tesla} from $-5$ to \SI{5}{\tesla} and back again.
For each magnetic field step we measured the reflected intensity multiple times in both $p$- and $s$-polarization.
In this case potential intensity fluctuations of the white light source are smoothed by normalizing each $p$-polarized spectrum to the spectrally integrated $s$-polarized intensity for each magnetic field step $I_p(B,\omega)/I_s(B)$.
Using the spectrally integrated $s$-polarized intensity for normalization has the advantage that otherwise small temperature deviations might slightly shift the optical resonances and consequently change the spectral dependence of TMOKE.

\printbibliography

\newpage

\noindent{\bf Acknowledgements}\\
This study was supported by the Deutsche Forschungsgemeinschaft via ICRC TRR 160 (Project C5), the Russian Foundation for Basic Research (project no. 16-32-60135 mol\textunderscore a\textunderscore dk) and the Foundation for the advancement of theoretical physics BASIS. The research in Poland was partially supported by the National Science Centre through the grant No. UMO-2017/25/B/ST3/02966, and by the Foundation of Polish Science through the IRA Programme co-financed by EU within SG OP.

\newpage
\begin{figure}
	\begin{center}
		\includegraphics[width=0.4\linewidth]{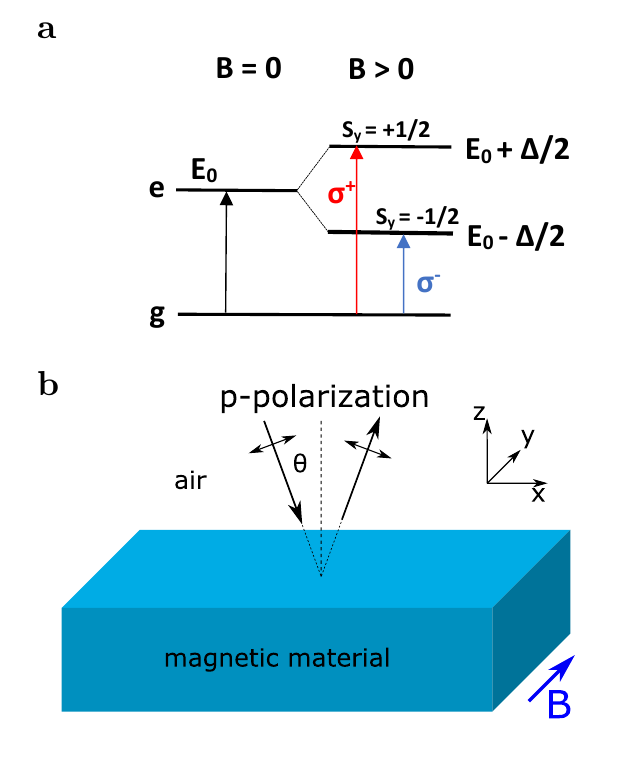}
		\caption{\textbf{a} Scheme of the optical transitions in a three-level system composed of the ground level (g) and two excited states (e), that are split off due to the Zeeman effect in magnetic field $\mathbf B \parallel Oy$. \textbf{b} The geometry of the considered setting, where $p$-polarized light is reflected from a semi-infinite magnetic medium with a three-level resonance shown in (\textbf{a}).}
		\label{Scheme}
	\end{center}
\end{figure}
\thispagestyle{empty}
\clearpage
\begin{figure}
	\begin{center}
		\includegraphics[width=\linewidth]{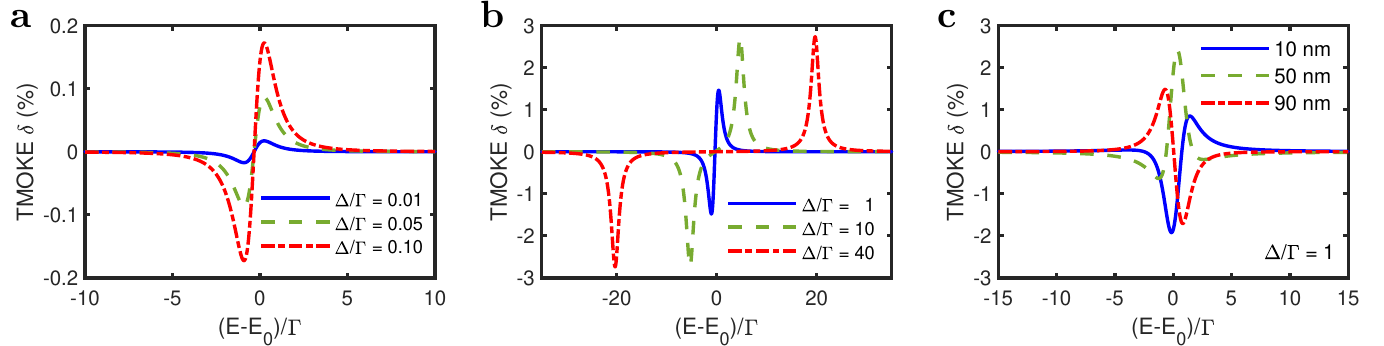}
		\caption{\textbf{a,b} TMOKE spectra for various $\Delta/\Gamma$ ratios, calculated by Eqs.~(\ref{deltaDef}) and (\ref{Eq:epsil}). Incidence angle $\theta=20^{\circ}$, $\Gamma=\SI{2}{\meV}$, $E_d=\SI{0.65}{\meV}$, $E_0=\SI{1.68}{\eV}$ and $\epsilon_b=10$. \textbf{c}~Calculated TMOKE spectra of the \SI{10}{\nano\meter}-thick magnetic film acting as a QW buried at three various distances from the sample surface. Incidence angle $\theta=20^{\circ}$, $\Gamma=\SI{2}{\meV}$, $E_d=\SI{0.65}{\meV}$, $\Delta=\SI{2}{\meV}$, $E_0=\SI{1.68}{\eV}$ and $\epsilon \approx 8$ for cap and epilayer. In all plots the dependencies on the relative change of energy with respect to the resonance linewidth are given.}
		\label{Profiles}
	\end{center}
\end{figure}
\thispagestyle{empty}
\clearpage
\begin{figure}
	\begin{center}
		\includegraphics[width=\linewidth]{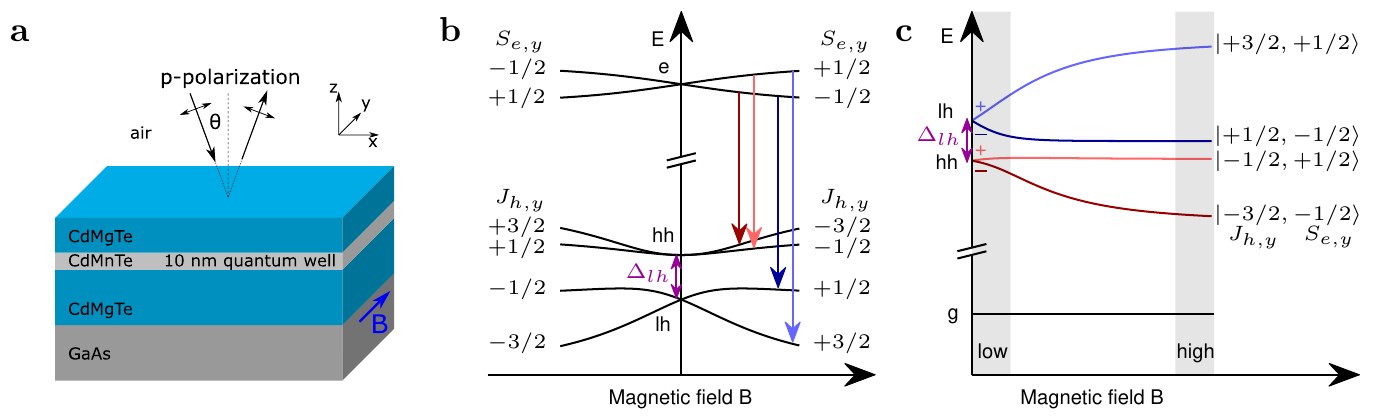}
		\caption{\textbf{a} Schematic presentation of the studied DMS QW structure and geometry of the experiment under light incidence. Magnetization $\textbf M$ and external magnetic field $\textbf B$ are oriented in the QW plane. \textbf{b} Energy diagram and Zeeman splitting of conduction and valence band states in Voigt geometry (single particle picture). Vertical lines indicate the relevant dipole-allowed optical transitions with elliptical polarization in $xz$-plane (optical transitions with linear polarization along magnetic field direction are not considered here). Red/blue lines correspond to $\sigma^+/\sigma^-$ polarization in the $xz$-plane in the limit of large magnetic fields where the electron and hole eigenstates are defined by the angular momentum projections on magnetic field direction $S_{e,y}$ and $J_{h,y}$, respectively. At $B=0$ the heavy-hole and light-hole states are split by $\Delta_{lh}$. \textbf{c} Energy diagram from (\textbf{b}) in the exciton picture. Gray areas indicate the limiting cases of low and high magnetic fields. For low magnetic fields the upper two transitions correspond to the light-hole exciton ($lh_{+,-}$, light and dark blue lines, respectively) while the lower ones correspond to the heavy-hole exciton ($hh_{+,-}$ light and dark red lines, respectively). Due to admixture from the light-holes to the heavy-holes the states at high magnetic field are denoted as labeled within the figure. The labeling of exciton states in this case is given by the $|J_{h,y},S_{e,y}\rangle$ notation.}
		\label{Fig:Sample}
	\end{center}
\end{figure}
\thispagestyle{empty}
\clearpage
\begin{figure}
	\begin{center}
		\includegraphics[width=\linewidth]{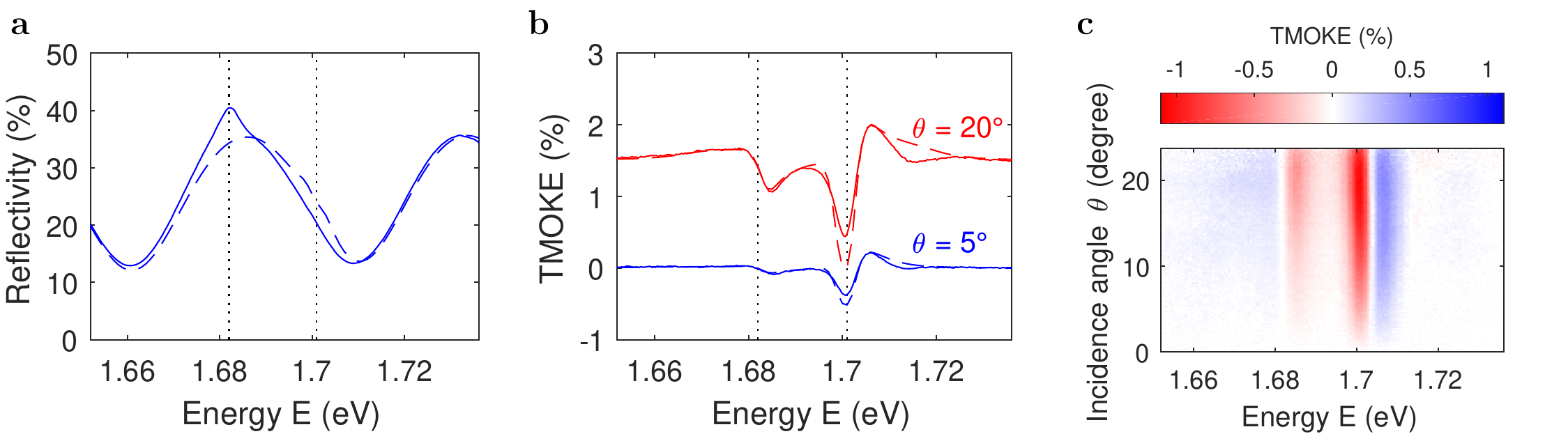}
		\caption{\textbf{a} Reflectivity spectrum from experiment (solid line) and theory (dashed line) at an incidence angle of $5^\circ$ and $T = \SI{10}{\K}$.
		\textbf{b} TMOKE spectra for $\theta = 5^\circ$ (blue) and $20^\circ$ (red) at $B = \SI{580}{\milli\tesla}$ and $T = \SI{10}{\K}$. The curves obtained for $20^\circ$ are shifted upwards by $1.5\%$ for clarity. Solid line refers to the experiment and dashed line corresponds to the theory.
		\textbf{c} Angular resolved TMOKE spectrum in the excitonic spectral range in the case of a weak coupling regime at $B = \SI{580}{\milli\tesla}$ and $T = \SI{10}{\K}$.
		$\Delta/\Gamma$ is different for light-hole and heavy-hole contributions, 1.79 for the light and 0.5 for the heavy one.}
		\label{Fig:Exp}
	\end{center}
\end{figure}
\thispagestyle{empty}
\clearpage
\begin{figure}
	\begin{center}
		\includegraphics[width=.5\linewidth]{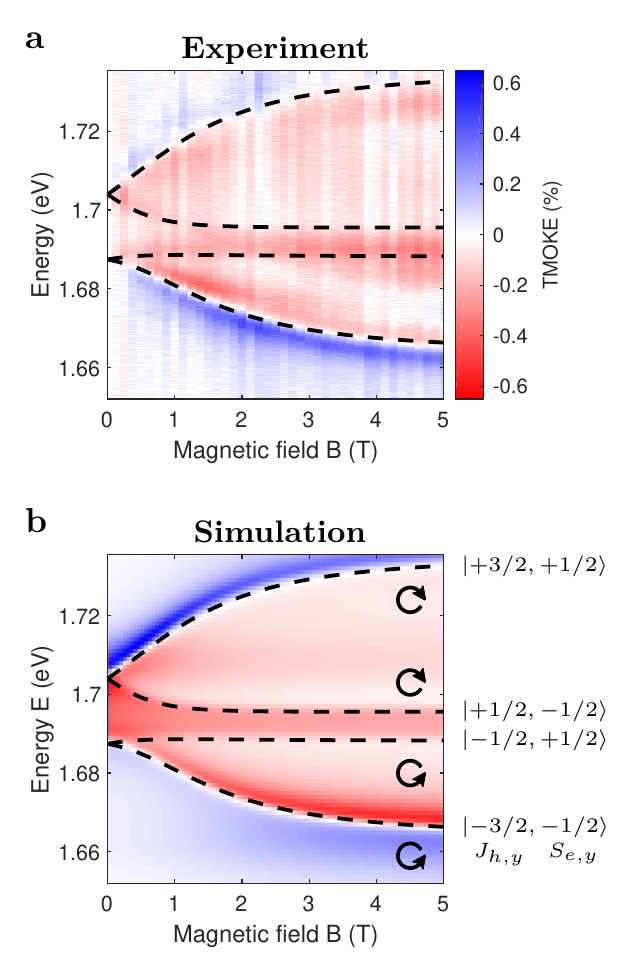}
		\caption{TMOKE in the excitonic spectral range in the strong coupling regime $\Delta/\Gamma > 1$.
		\textbf{a} Measured and \textbf{b} calculated energy and magnetic field resolved TMOKE spectra for $\theta = 5^\circ$ at $T=\SI{1.6}{\kelvin}$.}
		\label{Fig:HighField}
	\end{center}
\end{figure}
\thispagestyle{empty}
\clearpage
\setcounter{page}{1}
\renewcommand{\theequation}{S\arabic{equation}}
\setcounter{equation}{0}

\renewcommand{\thefigure}{S\arabic{figure}}
\setcounter{figure}{0}

{\setlength{\parindent}{0cm}
	\paragraph{Supplementary Information}$~~$\\
\textbf{A: TMOKE near resonance at the surface of semi-infinite material.}
Let us analyze TMOKE at the interface [air]/[magnetic material] near a resonance (see Fig.~\ref{Scheme}b). We consider a semi-infinite magnetic material with dielectric tensor $\epsilon$ given by Eq.~(\ref{Eq:epsil}). The analytical expression for TMOKE at the single interface of a semi-infinite medium is following~\cite{Zvezdin}
$$
\delta=-2 \Im \frac{4  \epsilon \tan \theta}{(\epsilon-1)(\epsilon-\tan \theta^2)} Q,
$$
where $\theta$ is the incidence angle. Light comes from the air and reflects at the magnetic dielectric. The magneto-optical response of the material is determined by $Q=g_y/\epsilon$. We substitute the expressions for the dielectric tensor elements from Eq.~(\ref{Eq:epsil}) and take into account that $4\pi |\mathbf{d}|$, $\tan \theta$ and $\epsilon_\text{b}$ are real
\begin{multline*}
\delta=- 16 \pi |\mathbf{d}|^2 \tan \theta \Im \left(\frac{\Delta}{\mathcal{D} (\epsilon_\text{b}(1+\frac{4\pi |\mathbf{d}|^2}{\epsilon_\text{b}} \frac{E_0- E - \rmi \Gamma}{\mathcal{D}})-1)}\times \right.\\
\left. \frac{1}{(\epsilon_\text{b}(1+\frac{4\pi |\mathbf{d}|^2}{\epsilon_\text{b}} \frac{E_0- E - \rmi \Gamma}{\mathcal{D}})-\tan^2 \theta)}\right),
\end{multline*}
where $\mathcal{D} = ( E_0-E - \rmi\Gamma )^2 - \Delta^2/4$.

Next, we assume that the incident angle $\theta$ is small enough so that we can neglect $\tan \theta^2$ with respect to $\epsilon_\text{b}$. For instance, in semiconductors the diagonal elements of dielectric tensor are about 10, therefore, this condition is satisfied up to incident angles of $\pi/3$; in case of ferromagnetic dielectrics with $\epsilon_\text{b} \approx 3$ this assumption is valid for $\theta \le \pi/4$. Therefore, the TMOKE has the following form
\begin{equation}
\delta=- 16 \pi |\mathbf{d}|^2 \tan \theta \Im \left(\frac{\Delta}{\mathcal{D} (\epsilon_\text{b}(1+\frac{4\pi |\mathbf{d}|^2}{\epsilon_\text{b}} \frac{E_0- E - \rmi\Gamma}{\mathcal{D}})-1)}\times \frac{1}{\epsilon_\text{b}(1+\frac{4\pi |\mathbf{d}|^2}{\epsilon_\text{b}} \frac{E_0- E - \rmi\Gamma}{\mathcal{D}})}\right).\label{eq:S1}
\end{equation}

One can see from Eq.~(\ref{eq:S1}) that the TMOKE depends linearly on the incidence angle $\theta$. The dependence on Zeeman splitting $\Delta$ is more complicated, so we perform an asymptotic analysis. We address TMOKE at small and large values of $\Delta$ with respect to the resonance half-linewidth $\Gamma$.

1. If $\Delta \ll \Gamma$ than $\mathcal{D} = ( E_0-E - \rmi\Gamma )^2 - \Delta^2/4 \approx ( E_0-E - \rmi\Gamma )^2$. With this assumption we rewrite Eq.~(\ref{eq:S1}) as
\begin{equation*}
\delta=-16 \pi |\mathbf{d}|^2 \tan \theta \Im \left(\frac{\Delta}{ (\epsilon_\text{b}-1)( E_0-E - \rmi\Gamma ) +4\pi |\mathbf{d}|^2} \times \frac{1}{\epsilon_\text{b}(E_0- E - \rmi\Gamma)+4\pi |\mathbf{d}|^2} \right).
\end{equation*}

One can see that $4 \pi |\mathbf{d}|^2/\epsilon_\text{b}$ should have the dimension of an energy, thus, we introduce the energy related to the dipole moment $E_d = 4 \pi |\mathbf{d}|^2/\epsilon_\text{b}$, so
\begin{equation*}
\delta=-4 E_d \epsilon_\text{b} \tan \theta \Im \left(\frac{\Delta}{ (\epsilon_\text{b}-1)( E_0-E - \rmi\Gamma ) +\epsilon_\text{b} E_d} \times \frac{1}{\epsilon_\text{b}(E_0- E - \rmi\Gamma)+\epsilon_\text{b} E_d} \right).
\end{equation*}

After simple calculations
\begin{equation*}
\delta=-\frac{ 8 E_d \Delta \Gamma \tan \theta}{ (\epsilon_\text{b}-1)} \frac{(E_0-E)+\frac{2\epsilon_\text{b}-1}{2(\epsilon_\text{b}-1)}E_d}{ (E_0-E+\frac{\epsilon_\text{b}}{\epsilon_\text{b}-1} E_d)^2 + \Gamma^2 } \times
\frac{1}{(E_0- E +E_d)^2+ \Gamma^2} .
\end{equation*}

When $\frac{2\epsilon_\text{b}-1}{2(\epsilon_\text{b}-1)} = 1+\frac{1}{2(\epsilon_\text{b}-1)}$ and $\frac{\epsilon_\text{b}}{\epsilon_\text{b}-1}=1+\frac{1}{\epsilon_\text{b}-1}$, then
\begin{equation*}
\delta=-\frac{ 8 E_d \Delta \Gamma \tan \theta}{(\epsilon_\text{b}-1)}\times
 \frac{E_0-E+E_d(1+\frac{1}{2(\epsilon_\text{b}-1)})}{ (E_0-E+(1+\frac{1}{\epsilon_\text{b}-1}) E_d)^2 + \Gamma^2 } \times \frac{1}{(E_0- E +E_d)^2+ \Gamma^2} .
\end{equation*}

We introduce the substitutions $\widetilde{E} = (E_0-E)+E_d(1+\frac{1}{2(\epsilon_\text{b}-1)})$ and $a=\frac{1}{2(\epsilon_\text{b}-1)}E_d$, and therefore we come to the following expression
\begin{align}
\begin{split}
\delta=-\frac{8 E_d \Delta \Gamma \tan \theta }{\epsilon_\text{b}-1}\frac{\widetilde{E}}{[ (\widetilde{E}+a)^2 + \Gamma^2][(\widetilde{E}-a)^2 + \Gamma^2 ]} .
\end{split}
\label{Eq:S2}
\end{align}

The denominator is the product of two quadratic functions of $\widetilde{E}$ that is symmetric with respect to $\widetilde{E}=0$. One can see that for $\Delta \ll \Gamma$ the TMOKE depends linearly on the Zeeman splitting $\Delta$, and the sign of TMOKE changes around the excitonic resonance energy.

Besides that, when $a \ll \Gamma$, the denominator approaches to $(\Gamma^2+\widetilde{E}^2)^2$ with the deviation of $a^2(2\Gamma^2-2\widetilde{E}^2+a^2)$. In this case the maximum deviation is estimated as $2a^2/\Gamma^2$ and Eq.~(\ref{Eq:S2}) is as follows
\begin{align}
\begin{split}
\delta=-\frac{8 E_d \Delta \Gamma \tan \theta }{\epsilon_\text{b}-1}\frac{\widetilde{E}}{(\widetilde{E}^2 + \Gamma^2)^2}.
\end{split}
\label{Eq:S3}
\end{align}

Thus, at $\widetilde{E}=\pm\Gamma/\sqrt{3}$ the TMOKE has extrema
\begin{align}
\begin{split}
\delta=\mp \frac{3 \sqrt{3} E_d \Delta \tan \theta }{2 ( \epsilon_\text{b}-1) \Gamma^2}.
\end{split}
\label{Eq:S4}
\end{align}

2. If $\Delta \gg \Gamma$ it should be taken into account that the resonances of TMOKE take place when $|E_0-E| = \Delta/2$. Let us consider the case $E_0-E = \Delta/2$. In this case $\mathcal{D} \approx - \rmi\Gamma \Delta- \Gamma^2$. Taking $\alpha=\Gamma/\Delta$ as a small parameter ($\alpha \to 0$, $\Delta \to \infty$) we obtain from Eq.~(\ref{eq:S1})
$$
\delta= \frac{4 \epsilon_\text{b} E_d \tan \theta}{\Gamma} \Im \left(\frac{1}{(\rmi+\alpha) (\epsilon_\text{b}-1 + E_d \frac{\rmi+\alpha}{2 \Gamma})(\epsilon_\text{b}+E_d\frac{\rmi+\alpha}{2 \Gamma})}\right)
$$

\begin{equation*}
\delta= \frac{4 \epsilon_\text{b} E_d \tan \theta}{\Gamma} \Im \left(\frac{1}{(\rmi+\alpha) }\times 
\frac{1}{\epsilon_\text{b}^2-\epsilon_\text{b}+(2\epsilon_\text{b}-1)E_d \frac{\rmi+\alpha}{2 \Gamma}+ E_d^2\frac{(\rmi+\alpha)^2}{4 \Gamma^2}}\right)
\end{equation*}

We calculate the imaginary part and neglect terms with $\alpha^2$
\begin{equation*}
\delta=- \frac{4 \epsilon_\text{b} E_d \tan \theta}{\Gamma} \times
\frac{\epsilon_\text{b}^2-\epsilon_\text{b}- \frac{E_d^2}{4\Gamma^2}+2\frac{2\epsilon_\text{b}-1}{2 \Gamma}E_d \alpha }{(\epsilon_\text{b}^2-\epsilon_\text{b}- \frac{E_d^2}{4\Gamma^2})^2+2\frac{2\epsilon_\text{b}-1}{2 \Gamma}E_d \alpha (\epsilon_\text{b}^2-\epsilon_\text{b}- \frac{E_d^2}{4\Gamma^2}) +(\frac{2\epsilon_\text{b}-1}{2 \Gamma})^2 E_d^2}\:.
\end{equation*}

The small parameter $\alpha$ is present in both nominator and denominator, thus we decompose the expression into a series and neglect terms with $\alpha$
\begin{align}
\begin{split}
\delta=- \frac{4 \epsilon_\text{b} E_d \tan \theta}{\Gamma} \frac{\epsilon_\text{b}^2-\epsilon_\text{b}- \frac{E_d^2}{4\Gamma^2}}{(\epsilon_\text{b}^2-\epsilon_\text{b}- \frac{E_d^2}{4\Gamma^2})^2+(\frac{2\epsilon_\text{b}-1}{2 \Gamma})^2 E_d^2}
\end{split}
\label{Eq:S5}
\end{align}

It can be easily shown that in case of $E_0-E = -\Delta/2$ the TMOKE will have the opposite sign than in Eq.~(\ref{Eq:S5}). Thus, when $\Delta \gg \Gamma$, the TMOKE is determined by the following equation
\begin{align}
\begin{split}
\delta=\mp \frac{4 \epsilon_\text{b} E_d \tan \theta}{\Gamma} \frac{\epsilon_\text{b}^2-\epsilon_\text{b}- \frac{E_d^2}{4\Gamma^2}}{(\epsilon_\text{b}^2-\epsilon_\text{b}- \frac{E_d^2}{4\Gamma^2})^2+(\frac{2\epsilon_\text{b}-1}{2 \Gamma})^2 E_d^2}
\end{split}
\label{Eq:S6}
\end{align}
The minus sign refers to $E_0-E>0$, and positive TMOKE corresponds to $E_0-E<0$.\\

\textbf{B: TMOKE at the heavy-hole and light-hole exciton resonances.}
Here, we describe the derivation of Eq.~\eqref{eq:chiSmall} and Eq.~\eqref{eq:chiBig} for the QW susceptibilities in Voigt geometry.

The scheme of the Zeeman splitting of the conduction and valence band states in the Voigt geometry
is shown in Fig.~\ref{Fig:Sample}b. The $\Gamma_{6}$ conduction band state splits into the states $\psi_{e,\uparrow(\downarrow)} = \uparrow(\downarrow)|s\rangle$ with the energies
\begin{equation}\label{eq:Ee}
E_{e,\pm 1/2}=E_{e}\pm 3{\cal A}\:,
\end{equation}
where $\mathcal{A}= \tfrac{1}{6} \alpha N_0 x S B_S(\epsilon)$, $|s\rangle$ is the $s$-symmetry Bloch amplitude and $\uparrow(\downarrow)$ are the spinors with $S_{e,y}=\pm 1/2$.
The exchange interaction parameters are introduced in the main text of Section \ref{TMOKE-lowB}.
$E_{e}$ is the energy of the electron ground state in the quantum well at zero magnetic field counted from the top of the bulk valence band.

The splitting of the $\Gamma_{8}$ valence band states is described by the following Luttinger Hamiltonian \cite{Landwehr,Winkler}
\begin{equation}
H_{\uparrow(\downarrow)}=\begin{pmatrix}
-(\gamma_{1}+\gamma_{2})k_{z}^{2}\pm 3\mathcal{B} &
\sqrt{3}\gamma_{2}k_{z}^{2}
\\\sqrt{3}\gamma_{2}k_{z}^{2}&-(\gamma_{1}-\gamma_{2})k_{z}^{2}\mp \mathcal B
\end{pmatrix}\:,\label{eq:HB}
\end{equation}
where the $\uparrow (\downarrow)$ subscripts label the subspaces of the Hamiltonian with $J_{h,y}=[3/2,-1/2]$ and $J_{h,y}=[-3/2,1/2]$ projections of the total angular momentum $\mathbf{J}_{h}$ on the magnetic field direction $y$.
Explicitly, the basis states read \cite{Ivchenko2005}:
\begin{align}\label{eq:basisH}
\psi_{h,3/2}&=-\frac{1}{\sqrt{2}}\uparrow (|Z\rangle+\rmi |X\rangle)\\\nonumber
\psi_{h,-3/2}&=\frac{1}{\sqrt{2}}\downarrow (|Z\rangle-\rmi |X\rangle)\\\nonumber
\psi_{h,1/2}&=\frac{\sqrt{6}|Y\rangle}{3}\uparrow-\frac{|Z\rangle+\rmi |X\rangle}{\sqrt{6}}\downarrow\\\nonumber
\psi_{h,-1/2}&=\frac{\sqrt{6}|Y\rangle}{3}\downarrow+\frac{|Z\rangle-\rmi |X\rangle}{\sqrt{6}}\downarrow\:,\nonumber
\end{align}
where $|X\rangle,|Y\rangle,|Z\rangle$ are the Bloch functions of corresponding symmetry and $\uparrow, \downarrow$ are the basis spinors. The wave vector $k_{z}\approx \pi/d$ in Eq.~\eqref{eq:HB} is the effective hole wave vector in $z$ direction ($d$ is the QW width). It phenomenologically describes the effect of size quantization assuming infinite QW walls.
The effect of strain is neglected and it is also assumed that excitons have zero in-plane wave vector.
The parameter $\mathcal B=1/6 N_0 x \beta S B_S(\epsilon)$ describes the exchange interaction with the Mn spins. We use the Luttinger parameters $\gamma_{1}\approx 5.3 \hbar^{2}/2m_{0}$ and $\gamma_{2}\approx 1.6 \hbar^{2}/2m_{0}$~\cite{Dang1982}, where $m_{0}$ is the free electron mass.

 The Hamiltonian Eq.~\eqref{eq:HB} describes the competition between the Zeeman splitting in the magnetic field $\mathbf{B}\parallel y$ and the size quantization along the $z$ axis. Its eigenstates have the following form:\cite{Winkler}
\begin{align}
E^{(h)}_{hh,-} &=-k_{z}^2\gamma_{1}-\mathcal{B}-2\sqrt{\mathcal{B}^2+\gamma_2^2k_{z}^4+\gamma_{2}k_{z}^2\mathcal{B}}\nonumber\:,\label{eq:Eh}\\
E^{(h)}_{hh,+} &=-k_{z}^2\gamma_{1}+\mathcal{B}-2\sqrt{\mathcal{B}^2+\gamma_2^2k_{z}^4-\gamma_{2}k_{z}^2\mathcal{B}}\nonumber\:,\\
E^{(h)}_{lh,-} &=-k_{z}^2\gamma_{1}-\mathcal{B}+2\sqrt{\mathcal{B}^2+\gamma_2^2k_{z}^4+\gamma_{2}k_{z}^2\mathcal{B}}\nonumber\:,\\
E^{(h)}_{lh,+} &=-k_{z}^2\gamma_{1}+\mathcal{B}+2\sqrt{\mathcal{B}^2+\gamma_2^2k_{z}^4-\gamma_{2} k_{z}^2\mathcal{B}}\:.
\end{align}
The spectrum is shown in Fig.~\ref{Fig:Sample}. In the case of large magnetic field the Zeeman splitting overcomes the light-heavy-hole splitting so that the size quantization effect can be neglected and the states are characterized by
the total momentum projection $J_{h,y}$ onto the field direction, as indicated in Fig.~\ref{Fig:Sample}.
The hole eigenstates are given directly by Eq.~\eqref{eq:basisH}.

At zero magnetic field the states with energies $E^{(h)}_{hh,\pm}$, $E^{(h)}_{lh,\pm}$ are degenerate and correspond to heavy and light-holes, respectively, with the splitting
\begin{equation}
\Delta_{lh}=|E^{(h)}_{lh,\pm}(B=0)-E^{(h)}_{hh,\pm}(B=0)|=4\gamma_{2}(\pi/d)^{2}.\label{eq:DeltaHL}
\end{equation}
An increase of the magnetic field leads to light-heavy-hole mixing.
At weak magnetic field the heavy-hole eigenstates can be written as
\begin{align}\label{eq:psihh0}
\psi_{hh,+}=\frac{\rmi}{\sqrt{2}}|X\rangle\uparrow&-\frac{1}{\sqrt{2}}|Y\rangle \downarrow+\\&\frac{\sqrt{2}}{12}\frac{\Delta_{h,F}}{\Delta_{lh}}(2|Z\rangle+\rmi |X\rangle)\uparrow\nonumber\\\nonumber
\psi_{hh,-}=\frac{\rmi}{\sqrt{2}}|X\rangle \downarrow&-\frac{1}{\sqrt{2}}|Y\rangle\uparrow+\\&\frac{\sqrt{2}}{12}\frac{\Delta_{h,F}}{\Delta_{lh}}(2|Z\rangle-\rmi |X\rangle)\downarrow\:.\nonumber
\end{align}
Here the main effect is the admixture of the $X$-type state from the light-holes to the heavy-holes which changes the symmetry of the heavy-hole states. This result is obtained by diagonalizing Eq.~\eqref{eq:HB} and substituting the Bloch functions from Eq.~\eqref{eq:basisH}. The parameter
\begin{equation}
\Delta_{h,F}=6\mathcal B
\end{equation}
is the giant Zeeman splitting of heavy-holes in Faraday geometry.
 The light-hole eigenstates at weak magnetic field read
\begin{align}\label{eq:psilh0}
\psi_{lh,+}=\frac{|Y\rangle}{\sqrt{6}}\downarrow\Bigl(1&+\tfrac{\Delta_{h,F}}{2\Delta_{lh}}\Bigr)+\\\nonumber&\frac{1}{\sqrt{6}}(2|Z\rangle+\rmi |X\rangle)\uparrow-\frac{\rmi\sqrt{6}}{12}\frac{\Delta_{hh,B}}{\Delta_{lh}}|X\rangle\uparrow\:.\\\nonumber
\psi_{lh,-}=\frac{|Y\rangle}{\sqrt{6}}\uparrow\Bigl(1&-\tfrac{\Delta_{h,F}}{2\Delta_{lh}}\Bigr)+\\\nonumber&\frac{1}{\sqrt{6}}(-2|Z\rangle+\rmi |X\rangle)\uparrow+\frac{\rmi\sqrt{6}}{12}\frac{\Delta_{h,F}}{\Delta_{lh}}|X\rangle\downarrow\:.\nonumber
\end{align}
Given the energy spectrum and the wave functions, we can calculate the effective permittivity tensor of the QW, neglecting the effect of spatial dispersion. It is given by the Kubo formula
\begin{equation}\label{eq:Kubo}
\eps_{mn}(\omega)=\eps_{b}\delta_{mn}+4\pi e^{2}(m_{0}/\omega)^{2}\times \sum\limits_{\nu=e,\pm 1/2}\sum\limits_{
\mu=hh\pm,lh\pm}\frac{\langle \mu|p_{m}|\nu\rangle \langle \nu|p_{n}|\mu\rangle}{\omega_{\nu}-\omega_{\mu}-\omega-\rmi \Gamma/\hbar}\:,
\end{equation}
where $\mathbf p$ is the momentum operator, $m_{0}$ is the free electron mass, and we keep only the resonant terms.
In the cubic approximation the only nonzero matrix elements are
\begin{equation}\label{eq:pcv}
p_{cv}=\langle S|p_{x}|X\rangle=\langle S|p_{y}|Y\rangle=\langle S|p_{z}|Z\rangle\:.
\end{equation}
It is convenient to introduce the LT splitting $\omega_{LT}$ as $4\pi e^{2}(m_{0}/\omega p_{cv})^{2}=\eps_{b}\omega_{LT}$.
Substituting Eqs.~\eqref{eq:psihh0},\eqref{eq:psilh0} or Eqs.~\eqref{eq:basisH} into Eq.~\eqref{eq:Kubo} we arrive to the permittivity tensors Eq.~\eqref{eq:chiSmall} and Eq.~\eqref{eq:chiBig} in the main text for weak and large magnetic fields, respectively. The optically allowed excitonic transitions have the energies
\begin{align}
E_{hh,+}&=E_{e,+1/2}-E^{(h)}_{hh,+}(B=0),\nonumber\\
E_{hh,-}&=E_{e,-1/2}-E^{(h)}_{hh,-}(B=0),\nonumber\\
E_{lh,+}&=E_{e,+1/2}-E^{(h)}_{lh,+}(B=0),\nonumber\\
E_{lh,-}&=E_{e,-1/2}-E^{(h)}_{lh,-}(B=0)\:.
\end{align}

\end{document}